\documentclass{article}
\usepackage{spconf,amsmath,graphicx}
\usepackage{subfig}

\title{Residual networks based DISTORTION CLASSIFICATION AND RANKING FOR LAPAROSCOPIC IMAGE QUALITY ASSESSMENT}

\name{Zohaib Amjad Khan$^{\star}$, Azeddine Beghdadi$^{\star}$, Mounir Kaaniche$^{\star}$ and Faouzi Alaya Cheikh$^{\dagger}$}

\address{
\begin{minipage}[b]{0.45\textwidth}\centering
\small{$^{\star}$ Universit\'e Sorbonne Paris Nord, \\
L2TI, UR 3043, F-93430, \\
Villetaneuse, France\\
zohaibamjad.khan@univ-paris13.fr \\
azeddine.beghdadi@univ-paris13.fr \\
mounir.kaaniche@univ-paris13.fr}
\end{minipage}
\begin{minipage}[b]{0.45\textwidth}\centering
\small{$^{\dagger}$  Norwegian University of Science and Technology (NTNU) \\ 
Norwegian Colour and Visual Computing Lab,\\
Gj{\o}vik,Norway, \\
faouzi.cheikh@ntnu.no}
\end{minipage}
}	
\begin{document}
%\ninept
%
\maketitle
\begin{abstract}
Laparoscopic images and videos are often affected by different types of distortion like noise, smoke, blur and non-uniform illumination. Automatic detection of these distortions, followed generally by application of appropriate image quality enhancement methods, is critical to avoid errors during surgery. In this context, a crucial step involves an objective assessment of the image quality, which is a two-fold problem requiring both the classification of the distortion type affecting the image and the estimation of the severity level of that distortion. Unlike existing image quality measures which focus mainly on estimating a quality score, we propose in this paper to formulate the image quality assessment task as a multi-label classification problem taking into account both the type as well as the severity level (or rank) of distortions. Here, this problem is then solved by resorting to a deep neural networks based approach. The obtained results on a laparoscopic image dataset show the efficiency of the proposed approach. 
\end{abstract}
\section{Introduction}
\label{sec:intro}

Image quality assessment (IQA) is a critical task \cite{wang2002image} in different vision-based systems like autonomous navigation, medical imaging, video-based surveillance etc \cite{wang2011applications}. In the medical field, there are various imaging modalities like CT, MRI, Ultrasound and laparoscopy where image quality assessment is very important both for analysis and diagnosis purposes. For instance, laparoscopic videos are often affected by different types of distortion like smoke, blur, noise and uneven illumination \cite{khan2020towards}. In this case, the knowledge of the kind of distortion affecting an image/video is necessary so that the appropriate quality enhancement method can be applied \cite{Sdiri2019,cong-cong2019}. This effectively makes IQA a two step process\cite{moorthy2010two}. In the first step, it is needed to detect the type of distortion affecting the image whereas in the second step the severity level of distortion needs to be quantified.

On the one hand, distortion classification is a challenging task and only few works have been developed to address it specifically. Indeed, many existing IQA methods like BIQI \cite{moorthy2010two}, DIIVINE \cite{moorthy2011blind} and BRISQUE \cite{mittal2012no} employ a distortion classification stage only to get the class probabilities for an image, which are then used as weights for the results (i.e scores) of the regression step. Although their classification output could be used separately, yet the classification in all these methods is based on non-generic and hand-crafted features. Hence, there is a big gap when it comes to more generic distortion classification for quality monitoring tasks. In this paper, one of the aims is to address this challenge in the case of laparoscopic videos.

On the other hand, the problem of evaluating the severity of distortions in an image is often addressed in the form of an absolute quality score. Most of the state-of-the-art methods are based on extracting specific features from the images followed by supervised learning using subjective scores to get a quality score \cite{ShahkolaeiBC19}. Such methods have two disadvantages. First, they strongly depend on the specific distortion categories and the type of distribution used for coefficients modeling to extract features. For this reason, they may not be well adapted to other distortions like those seen in laparoscopic images. Secondly, the subjective scores used for learning are usually inconsistent and may be biased by image content \cite{gao2015learning}. Additionally, using methods dependent on subjective scores is not usually possible with medical data, like laparoscopic images, due to the limited image data and the lack of public quality datasets. Moreover, while non-medical quality evaluations are primarily aimed at recording perceptual or aesthetic quality, medical images need diagnostic quality evaluations by medical experts, which is a non-trivial task.

To overcome the problems of lack of labeled data and inconsistencies in subjective scoring, learning based ranking approaches for quality assessment have recently been developed \cite{liu2017rankiqa}. These approaches predict the distortion severity level instead of the absolute quality score and hence do not require subjective scores. The data for these methods can also be easily generated since the ranks are assigned during the distorted data generation process. A similar ranking-based approach has also been used for contrast enhancement evaluation \cite{ChetouaniQDB19}. Hence, in this paper, we have also chosen the ranking method for estimation of severity level during IQA stage for laparoscopic images. 

In light of the above studies, we propose here a new methodology for IQA of laparoscopic images in the context of quality monitoring applications, whereby our focus is on the two important problems of distortion classification and quality ranking. More precisely, we propose to solve these two problems using a single deep neural network based approach. In order to achieve this, we have combined these two sub-problems to formulate them as a single multi-label classification problem and then applied residual networks to perform distortion classification and ranking simultaneously. 

The remainder of this paper is organized as follows. Section II gives an overview of the related works. This is followed by Section III which presents our methodology to solve the multi-label classification problem. In Section IV, the experimental results are provided and discussed. Finally, some conclusions and future work are given in Section V.

\section{Related works}
\label{sec:related}
The use of hand-crafted features for IQA has the limitation of being application specific and non-generic. Convolutional neural networks (CNN) allow to overcome this issue by achieving an automated feature learning approach that can be extended to all kinds of distortions. CNNs have been shown to be very effective in several image processing tasks including object classification, enhancement, segmentation and retrieval. For this reason, the use of different deep neural networks (DNN) based on CNN have also been proposed for image quality assessment.  

However, the use of deep learning approaches for IQA is not a straightforward task and there are many variations for its application to IQA. For instance, in \cite{varga2018deeprn}, a blind IQA is proposed based on DNN to predict distribution of quality rating instead of the scores. They have used ResNet-101 with Huber's loss function using full size images as input. Similarly, in \cite{talebi2018nima}, the authors have explored VGG16, Inception-v2 and MobileNet to also predict distribution of aesthetic and technical quality ratings using Earth Mover's Distance loss function. Moreover, Ma \textit{et al.} \cite{ma2017end} have tackled IQA using a multi-task DNN using two sub-networks with shared initial layers. They have proposed to use one of the sub-networks for distortion identification and the other one for quality prediction. The output of the first sub-network is a class probability vector which is then combined with the output of the second sub-network in a weighted summation. Their approach is close to the proposed one which presents the advantage of resorting to one single opinion-unaware network for both distortion classification and ranking.

Besides these methods, some CNN-based methods for IQA also take image patches as inputs \cite{kang2014convolutional, bosse2016deep, chetouani2018blind}. Since the subjective score exists for the entire image, these methods either assume total score as an average of all patches or as a weighted average, based on saliency or output of another sub-network. Some recent works have also proposed techniques for quality based image ranking. For instance, Xu \textit{et al.} \cite{xu2016multi} use machine learning based approach to predict ranking for images before evaluating the absolute score. More recently, in \cite{liu2017rankiqa}, Siamese neural network has been employed to learn ranks of images. However, unlike our proposed approach here, they are only focused on the aspect of quality ranking and do not address the distortion classification problem.  

\vspace{-0.25cm}
\section{Proposed multi-label distortion classification and ranking with ResNet}
\label{sec:methodology}
In this section, we describe the proposed methodology for laparoscopic image quality assessment. As mentioned earlier, the distortion classification and quality ranking tasks are jointly performed using a single multi-label classification problem. Indeed, a multi-label classification problem can be transformed into a single-label multiclass classification task using one of the Problem Transformation methods \cite{tsoumakas2007multi}. Among these methods, Label Powerset approach is simpler and works well for a smaller number of total labels. In this approach, each combination of labels in the dataset is considered as a separate class resulting in $C$ classes in total. This task can then be solved using an appropriate multiclass classification method.

\subsection{Multi-label classification formulation}
More precisely,  given a set of $N$ possible distortions, $D = \{D_1,D_2,...,D_N\}$ and another set of $M$ possible discernible distortion levels, $L = \{L_1,L_2,...,L_M\}$, each image can be labeled with one label $D_{i}L_{j}$, where $1 \leq i \leq N$ and $1 \leq j \leq M$. For our laparoscopic data $X$, we have chosen five different distortion classes that often affect laparoscopic images. These are defocus blur (DB), motion blur (MB), additive white Gaussian noise (AWGN), smoke (SM) and uneven illumination (UI). For distortion severity, we have considered four levels, namely  hardly visible (HV), just noticable (JN), visible and annoying (VA) and extremely annoying (EA). Hence, with these labels we have the following two sets:
\begin{align}
& D_X = \{DB,MB,WN,SM,UI\} \\
&L_X = \{HV,JN,VA,EA\}
\end{align}
We have opted for discrete semantic severity levels instead of continuous in light of their ease of use and understanding in clinical context. For notations concision, the labels indicating the distortion level will be ranked from 1 to 4 while 4 being the worst (EA). After the multiclass reformulation step, we get the single following set of $C=20$ classes:
\begin{align}
C_X = \{&(DB_j)_{1\leq j \leq 4}, (MB_j)_{1\leq j \leq 4}, (WN_j)_{1\leq j \leq 4}, \nonumber \\ & (SM_j)_{1\leq j \leq 4}, (UI_j)_{1\leq j \leq 4}\}
\end{align}

Before solving this multi-label classification problem, let us first describe the proposed approaches used to simulate the 5 typical distortions considered above.

\subsection{Distortions generation}
For reference images, we have extracted different frames from the videos in Cholec80 dataset \cite{twinanda2017endonet}. The latter contains 80 different videos of cholecystectomy surgeries. For our training and test dataset, we have tried to select images with an attempt to include maximum possible variations of the scene content. Examples of these variations are images containing content like organs (liver and gall-bladder), vessels, blood, single and multiple instruments, water, staples, cut and stretched tissue and organ extraction bag. 

\begin{figure}[!h]
	\centering
	\subfloat[$MB_1$ \label{fig:mb1}]{\includegraphics[width=0.1\textwidth]{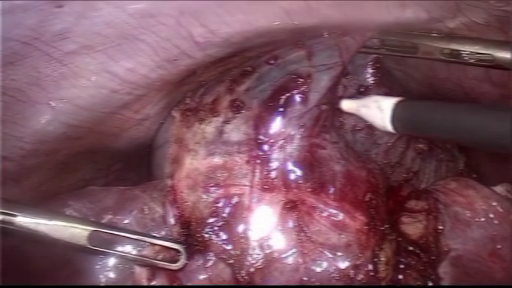}}\hspace{0.01\textwidth}
	\subfloat[$MB_4$ \label{fig:mb4}] {\includegraphics[width=0.1\textwidth]{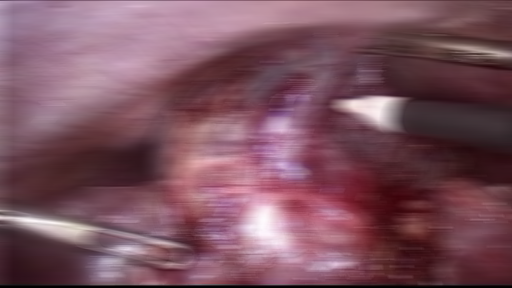}}\hspace{0.01\textwidth}
	\subfloat[$DB_1$ \label{fig:df1}]{\includegraphics[width=0.1\textwidth]{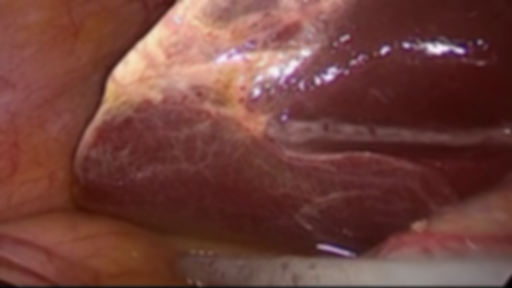}}\hspace{0.01\textwidth}
	\subfloat[$DB_4$ \label{fig:df4}]{\includegraphics[width=0.1\textwidth]{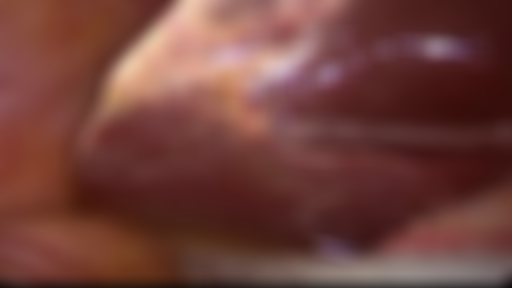}}\hspace{0.01\textwidth} \\ \vspace{0.15cm}
	\subfloat[$UI_1$\label{fig:ui1}] {\includegraphics[width=0.1\textwidth]{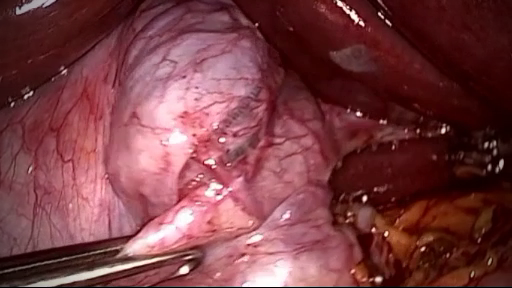}}\hspace{0.01\textwidth}
	\subfloat[$UI_4$\label{fig:ui4}]{\includegraphics[width=0.1\textwidth]{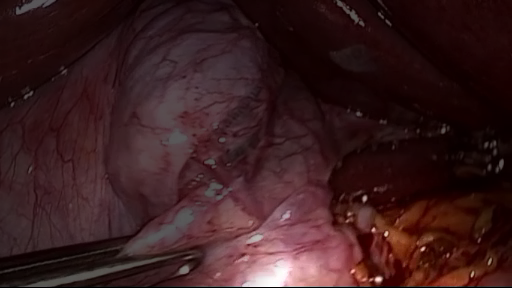}}\hspace{0.01\textwidth}
	\subfloat[$SM_1$ \label{fig:sm1}]{\includegraphics[width=0.1\textwidth]{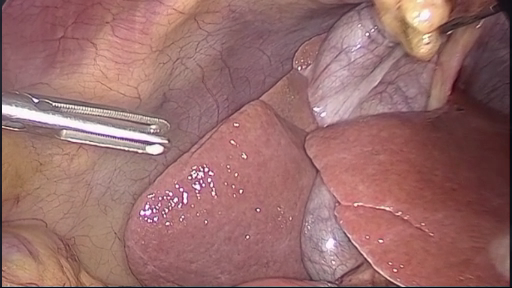}}\hspace{0.01\textwidth}
	\subfloat[$SM_4$\label{fig:sm4}] {\includegraphics[width=0.1\textwidth]{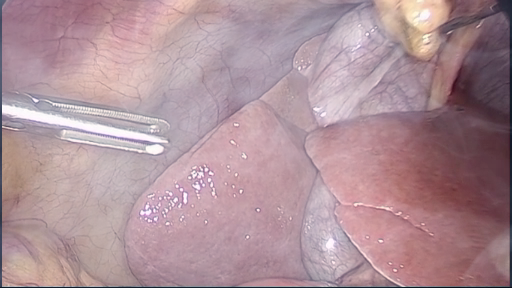}}\hspace{0.01\textwidth} \\ \vspace{0.15cm}
	\subfloat[$WN_1$ \label{fig:noise1}]{\includegraphics[width=0.1\textwidth]{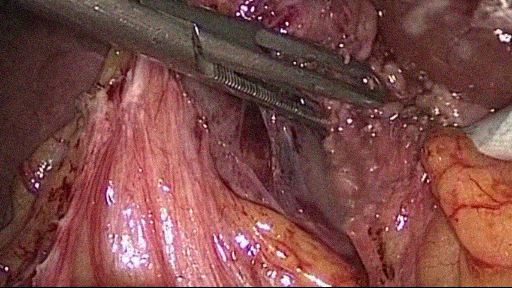}}\hspace{0.01\textwidth}
	\subfloat[$WN_4$ \label{fig:noise4}]{\includegraphics[width=0.1\textwidth]{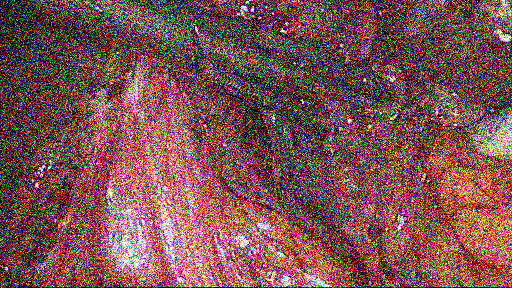}}\hspace{0.01\textwidth}
	\caption{\label{fig:diff_classes} Generated distorted images at levels 1 and 4 for different samples of our laparoscopic dataset.} 
	\label{fig:dist_frames}
	\vspace{-0.3cm}
\end{figure}

To generate synthetic distortions in the selected reference images, we have used MATLAB to simulate all the distortions with different levels for our database. For \textit{defocus blur}, a symmetric low-pass Gaussian filter was applied to each image. The filter size and the standard deviation were changed to generate different levels of this distortion. For directional and uniform motion blur, due to camera or/and object motion during the image acquisition, a linear filter was used with various values of the kernel motion parameter to generate different blurring levels . Similarly, for \textit{noise} distortion  an additive white Gaussian noise was generated. In this case, variance of the Gaussian distribution was adjusted to generate multiple levels of noisy images. For \textit{uneven illumination}, a special gray-scale mask was generated having a circular bright region of high intensity values. The areas surrounding the circular region were generated with decreasing intensities as a function of distance from the center of the bright region. The multiplication of this mask with the original image gave us the unevenly illuminated distorted image. By changing the two parameters of the center location of the bright region and its area, we generated four different levels for uneven illumination. Finally, in order to generate \textit{smoke}, we have used screen blending model \cite{khan2020towards}. In this technique, real smoke image having a black background is combined with the reference image in such a way that black areas produce no change to the original image while the brighter areas overlay the original ones. Using different opacity levels for the smoke frame we have generated four different levels of severity for smoke. An illustration of the generated distorted images at levels 1 and 4 is provided in Fig. \ref{fig:dist_frames} for different samples of our laparoscopic dataset. 
Once the distortion generation process as well as the multi-label classification problem are described, the latter will be solved by using a deep learning approach based on the residual network architecture.
\subsection{Residual network architecture}
\vspace{-0.2cm}
Deeper convolutional neural networks, which simply stacked many layers that exceed a certain threshold may result in gradient vanishing problem and lead to an accuracy drop \cite{he2016deep}. To alleviate this problem, residual network, which replaces direct stacked layers by skip connections that sum-up the output of a block of layers to its input, has been developed \cite{he2016deep}. Residual network, referred to as ResNet, is considered as a class of deep neural networks which is commonly applied to image recognition as well as classification \cite{he2016deep}. For instance, ResNet-50 is a deep network, with 50 layers and over 23 million trainable parameters, which has excellent generalization performance. For these reasons, we have retained the ResNet architecture, with different numbers of layers, to achieve our distortion classification and ranking tasks. To this end, for a given ResNet architecture with a fixed number of layers, only the output fully-connected layer is replaced with a layer of size 20 corresponding to the number of classes in the dataset. Then, the training of the model is done on our laparoscopic dataset. This is performed by using the cross-entropy as a loss function, which is often considered in classification problems. This function is defined as follows:
\begin{equation}
\label{cross_ent}
\mathcal{L}=-\sum\limits_{c=1}^{C}y_{o,c}\log(p_{o,c})
\end{equation}
where $C=20$ is the number of classes in our distorted laparoscopic dataset,  $y_{o,c}$ is a binary indicator with value 1 if the class label $c$ is the correct classification for the observation $o$, and 0 if not, and $ p_{o,c} $ is the network predicted probability that the observation $o$ is of class $c$. The minimization of the above loss function is achieved using the Adam optimizer.

\section{Experimental Results}

\begin{table}[!h]
	\begin{center}
		\begin{tabular}{|l|l|l|}
			\hline
			\textbf{Hyperparameters}& \textbf{RankIQA} & \textbf{Proposed}\\ 
			\hline
			\textbf{Learning rate} & 1e-5 & 1e-3 \\ 
			\hline					
			\textbf{Epochs} & 100 & 300 \\	
			\hline	
			\textbf{Mini-batch size} & 5 & 10 \\	
			\hline
			\textbf{Loss function} & Hinge loss & Cross-entropy\\	
			\hline
			\textbf{Optimizer} & SGD & Adam\\	
			\hline
		\end{tabular}	
	\end{center}
	\caption{Hyperparameters for the two networks used for comparison}
	\label{hyper}
	\vspace{-0.2cm}
\end{table}

To evaluate the performance of the proposed methodology, we have used 1000 images of size $512 \times 288$ extracted from different videos of Cholec80 laparoscopic dataset \cite{twinanda2017endonet}. As mentioned in Section 3, we selected images with various contents like organs (liver and gall-bladder), vessels, blood, single and multiple instruments, etc. Moreover, five distortions at four different severity levels were then generated for all the images resulting in a total of 20000 images (1000 images per class). Since the images were distorted synthetically, we already had the labels for all images corresponding to distortion type and severity level and hence did not need the input of clinical experts. Eighty percent of image data from each class (800 images) was used for training of ResNet while the remaining twenty percent of images were reserved for testing. The training and testing for ResNet model used to solve our multi-label classification problem was carried out by using PyTorch framework on an NVIDIA GeForce GTX 1080, 64 Gb GPU. The learning rate used to train our model was set to 0.001, the mini-batches of size was set to 10, and the number of epochs was set to 300 (see Table~\ref{hyper}). 

To the best of our knowledge, our methodology is the first  one which aims to perform distortion classification as well as ranking simultaneously. For this reason, we will propose to compare our approach separately to classification methods and ranking methods. Indeed, the results of quality ranking from our method are firstly compared to a recent deep learning based method which employs a Siamese network for ranking before fine-tuning its single trained branch to obtain quality scores \cite{liu2017rankiqa}. The latter is designated by RankIQA. For our comparison, we have only used the ranking part of their network with Hinge loss (see Table~\ref{hyper}). In order to evaluate these ranking methods, we have used Spearman Rank-Order Correlation Coefficient (SROCC). Table~\ref{srocc} provides the SROCC values for the RankIQA method as well as the proposed one using ResNet-18. The obtained results show that our approach yields better ranking performance compared to the recent deep learning-based RankIQA method.

\begin{table}[tp]
	\begin{center}
		\begin{tabular}{|l|l|}
			\hline
			\textbf{Methods}& \textbf{SROCC}\\ 
			\hline
			\textbf{Proposed one (with ResNet-18)}& \textbf{0.69} \\ 
			\hline					
			\textbf{RankIQA \cite{liu2017rankiqa}}& 0.57 \\	
			\hline	
		\end{tabular}	
	\end{center}
	\caption{SROCC of ranking methods with laparoscopic dataset}
	\label{srocc}
\end{table}

\begin{table}[tp]
	\begin{center}
		\begin{tabular}{|l|l|}
			\hline
			\textbf{Proposed method with}& \textbf{Accuracy}\\ 
			\hline
			 \textbf{ResNet-18}& 83.3\%\\ 
			\hline					
			\textbf{ResNet-34}& 84.7\% \\	
			\hline
			\textbf{ResNet-50}& \textbf{87.3}\% \\	
			\hline	
		%	\textbf{BRISQUE}& 99.0\% \\	
		%	\hline
		%	\textbf{BIQI}& 98.4\% \\	
	    %		\hline	
		\end{tabular}	
	\end{center}
	\caption{Classification accuracy of the proposed method with different ResNet models for our laparoscopic dataset.}
	\label{classification_acc}
\end{table}

Moreover, the performance of the proposed methodology is evaluated in terms of classification accuracy. To this end, and in addition to ResNet-18, two other deeper variants with more layers have been considered. Table~\ref{classification_acc} shows the mean classification accuracy, for ResNet-18, ResNet-34 and ResNet-50, after 10 repetitions. Thus, it can be observed that the used ResNet-18 leads to good accuracy results (around 83\%). Moreover, by using deeper ResNet architectures, the accuracy is improved and reaches 87.3\% with ResNet-50. It is worth pointing out that classical machine learning approaches, like BRISQUE and BIQI, as well as deep learning approaches have not been considered in this classification accuracy comparison since the latter perform generally classification based \textit{only} on the distortion type (i.e. 5 classes corresponding to the five kinds of distortion) whereas our methodology uses 20 classes as it jointly performs distortion classification and its severity ranking. 

Therefore, the obtained results show the benefits of the proposed approach for efficient quality ranking of the distortion severity level while yielding good classification accuracy. 

\vspace{-0.3cm}
\section{Conclusion and perspectives}
In this paper, we have proposed a novel strategy of image quality assessment for quality monitoring tasks especially in the context of laparoscopic images. More precisely, a single neural network (ResNet) is used to solve the two most important IQA sub-tasks, namely distortion classification and quality ranking, thanks to a multi-label classification formulation. Then, we have tested our method on a laparoscopic dataset and obtained promising results. In future, we plan to consider other kinds of distortions including distortion mixtures and an additional step of quality score prediction using a labeled laparoscopic image dataset with human subjective scores. 
\vspace{-0.3cm}
\section{Acknowledgment}
This research work is part of a project that has received funding from the European Union's Horizon 2020 research and innovation programme under grant agreement No 722068.

\bibliographystyle{IEEEbib}
\bibliography{refs}

\begin{thebibliography}{10}

\bibitem{wang2002image}
Z.~Wang, A.~C. Bovik, and L.~Lu,
\newblock ``Why is image quality assessment so difficult?,''
\newblock in {\em International Conference on Acoustics, Speech, and Signal
  Processing}. IEEE, 2002, vol.~4, pp. 3313--3316.

\bibitem{wang2011applications}
Z.~Wang,
\newblock ``Applications of objective image quality assessment methods
  [applications corner],''
\newblock {\em IEEE Signal Processing Magazine}, vol. 28, no. 6, pp. 137--142,
  2011.

\bibitem{khan2020towards}
Z.~A. Khan, A.~Beghdadi, F.~A. Cheikh, M.~Kaaniche, E.~Pelanis, R.~Palomar,
  {\AA}.~A. Fretland, B.~Edwin, and O.~J. Elle,
\newblock ``Towards a video quality assessment based framework for enhancement
  of laparoscopic videos,''
\newblock in {\em Medical Imaging 2020: Image Perception, Observer Performance,
  and Technology Assessment}. International Society for Optics and Photonics,
  2020, vol. 11316, p. 113160P.

\bibitem{Sdiri2019}
B.~Sdiri, M.~Kaaniche, F.~Alaya-Cheikh, A.~Beghdadi, and O.~J. Elle,
\newblock ``Efficient enhancement of stereo endoscopic images based on joint
  wavelet decomposition and binocular combination,''
\newblock {\em IEEE Transactions on Medical Imaging}, vol. 38, no. 1, pp.
  33--45, 2019.

\bibitem{cong-cong2019}
C.~Wang, A.~K. Mohammed, F.~A. Cheikh, A.~Beghdadi, and O.~J. Elle,
\newblock ``Multiscale deep desmoking for laparoscopic surgery,''
\newblock in {\em Medical Imaging 2019: Image Processing, San Diego,
  California, United States, 16-21 February 2019}, 2019, p. 109491Y.

\bibitem{moorthy2010two}
A.~K. Moorthy and A.~C. Bovik,
\newblock ``A two-step framework for constructing blind image quality
  indices,''
\newblock {\em IEEE Signal processing letters}, vol. 17, no. 5, pp. 513--516,
  2010.

\bibitem{moorthy2011blind}
A.~K. Moorthy and A.~C. Bovik,
\newblock ``Blind image quality assessment: From natural scene statistics to
  perceptual quality,''
\newblock {\em IEEE transactions on Image Processing}, vol. 20, no. 12, pp.
  3350--3364, 2011.

\bibitem{mittal2012no}
A.~Mittal, A.~K. Moorthy, and A.~C. Bovik,
\newblock ``No-reference image quality assessment in the spatial domain,''
\newblock {\em IEEE Transactions on image processing}, vol. 21, no. 12, pp.
  4695--4708, 2012.

\bibitem{ShahkolaeiBC19}
A.~Shahkolaei, A.~Beghdadi, and M.~Cheriet,
\newblock ``Blind quality assessment metric and degradation classification for
  degraded document images,''
\newblock {\em Sig. Proc.: Image Comm.}, vol. 76, pp. 11--21, 2019.

\bibitem{gao2015learning}
F.~Gao, D.~Tao, X.~Gao, and X.~Li,
\newblock ``Learning to rank for blind image quality assessment,''
\newblock {\em IEEE transactions on neural networks and learning systems}, vol.
  26, no. 10, pp. 2275--2290, 2015.

\bibitem{liu2017rankiqa}
X.~Liu, J.~van~de Weijer, and A.~D. Bagdanov,
\newblock ``Rankiqa: Learning from rankings for no-reference image quality
  assessment,''
\newblock in {\em Proceedings of the IEEE International Conference on Computer
  Vision}, 2017, pp. 1040--1049.

\bibitem{ChetouaniQDB19}
A.~Chetouani, M.~Ali. Qureshi, M.~A. Deriche, and A.~Beghdadi,
\newblock ``A novel ranking algorithm of enhanced images using a convolutional
  neural network and a saliency-based patch selection scheme,''
\newblock in {\em 11th International Conference on Quality of Multimedia
  Experience}, Berlin, Germany, June 2019, pp. 1--6.

\bibitem{varga2018deeprn}
D.~Varga, D.~Saupe, and T.~Szir{\'a}nyi,
\newblock ``Deeprn: A content preserving deep architecture for blind image
  quality assessment,''
\newblock in {\em International Conference on Multimedia and Expo}. IEEE, 2018,
  pp. 1--6.

\bibitem{talebi2018nima}
H.~Talebi and P.~Milanfar,
\newblock ``Nima: Neural image assessment,''
\newblock {\em IEEE Transactions on Image Processing}, vol. 27, no. 8, pp.
  3998--4011, 2018.

\bibitem{ma2017end}
K.~Ma, W.~Liu, K.~Zhang, Z.~Duanmu, Z.~Wang, and W.~Zuo,
\newblock ``End-to-end blind image quality assessment using deep neural
  networks,''
\newblock {\em IEEE Transactions on Image Processing}, vol. 27, no. 3, pp.
  1202--1213, 2017.

\bibitem{kang2014convolutional}
L.~Kang, P.~Ye, Y.~Li, and D.~Doermann,
\newblock ``Convolutional neural networks for no-reference image quality
  assessment,''
\newblock in {\em Proceedings of the IEEE conference on computer vision and
  pattern recognition}, 2014, pp. 1733--1740.

\bibitem{bosse2016deep}
S.~Bosse, D.~Maniry, T.~Wiegand, and W.~Samek,
\newblock ``A deep neural network for image quality assessment,''
\newblock in {\em International Conference on Image Processing}. IEEE, 2016,
  pp. 3773--3777.

\bibitem{chetouani2018blind}
A.~Chetouani,
\newblock ``A blind image quality metric using a selection of relevant patches
  based on convolutional neural network,''
\newblock in {\em European Signal Processing Conference}. IEEE, 2018, pp.
  1452--1456.

\bibitem{xu2016multi}
L.~Xu, J.~Li, W.~Lin, Y.~Zhang, L.~Ma, Y.~Fang, and Y.~Yan,
\newblock ``Multi-task rank learning for image quality assessment,''
\newblock {\em IEEE Transactions on Circuits and Systems for Video Technology},
  vol. 27, no. 9, pp. 1833--1843, 2016.

\bibitem{tsoumakas2007multi}
G.~Tsoumakas and I.~Katakis,
\newblock ``Multi-label classification: An overview,''
\newblock {\em International Journal of Data Warehousing and Mining (IJDWM)},
  vol. 3, no. 3, pp. 1--13, 2007.

\bibitem{twinanda2017endonet}
A.~P. Twinanda, S.~Shehata, D.~Mutter, J.~Marescaux, M.~De~Mathelin, and
  N.~Padoy,
\newblock ``Endonet: a deep architecture for recognition tasks on laparoscopic
  videos,''
\newblock {\em IEEE transactions on medical imaging}, vol. 36, no. 1, pp.
  86--97, 2017.

\bibitem{he2016deep}
K.~He, X.~Zhang, S.~Ren, and J.~Sun,
\newblock ``Deep residual learning for image recognition,''
\newblock in {\em Proceedings of the IEEE conference on computer vision and
  pattern recognition}, 2016, pp. 770--778.

\end{thebibliography}

\end{document}